\documentclass[10pt,journal,compsoc]{IEEEtran}

\ifCLASSOPTIONcompsoc
\usepackage[nocompress]{cite}
\else
\usepackage{cite}
\fi

\usepackage{booktabs}
\usepackage{multirow}
\usepackage{color}
\usepackage{xcolor}
\usepackage{subfigure}
\newcommand{\etal}{\textit{et al.}}
\ifCLASSINFOpdf
\usepackage[pdftex]{graphicx}
\else
\usepackage[dvips]{graphicx}
\fi
\usepackage{amsmath}
\usepackage{bm}

\begin{document}

	\title{PRS-Net: Planar Reflective Symmetry Detection Net for 3D Models}

	\author{
		Lin Gao, Ling-Xiao Zhang, Hsien-Yu Meng, Yi-Hui Ren, Yu-Kun Lai, Leif Kobbelt
		\IEEEcompsocitemizethanks{
			\IEEEcompsocthanksitem L. Gao, L.X. Zhang and Y.H. Ren are with Institute of Computing Technology, Chinese Academy of Sciences, Beijing, China.
			\protect\\
			E-mail:\{gaolin, zhanglingxiao\}@ict.ac.cn, renyihui17@mails.ucas.ac.cn
			\IEEEcompsocthanksitem H.Y. Meng is with University of Maryland, College Park, Maryland, US.
			\protect\\
			E-mail:mengxy19@cs.umd.edu
			\IEEEcompsocthanksitem Y.K. Lai is with Visual Computing Group, School of Computer Science and Informatics, Cardiff University, Wales, UK.
			\protect\\
			E-mail:LaiY4@cardiff.ac.uk
			\IEEEcompsocthanksitem L. Kobbelt is with Institute for Computer Graphics and Multimedia, RWTH Aachen University, Aachen, Germany.
			\protect\\
			E-mail:kobbelt@cs.rwth-aachen.de

		}
		\thanks{Manuscript received April 19, 2019; revised August 26, 2019.}}

	\markboth{IEEE TRANSACTIONS ON VISUALIZATION AND COMPUTER GRAPHICS,~Vol.~xx, No.~xx, July~2019}%
	{Lin \MakeLowercase{\textit{et al.}}: PRS-Net: Planar Reflective Symmetry Detection Net for 3D Models}

	\IEEEtitleabstractindextext{%
		\begin{abstract}
			In geometry processing, symmetry is a universal type of high-level structural information of 3D models and benefits many geometry processing tasks including shape segmentation, alignment,  matching, and completion. Thus it is an important problem to analyze various symmetry forms of 3D shapes. Planar reflective symmetry is the most fundamental one. Traditional methods based on spatial sampling can be time-consuming and may not be able to identify all the symmetry planes. In this paper, we present a novel learning framework to automatically discover global planar reflective symmetry of a 3D shape. Our framework trains an unsupervised 3D convolutional neural network to extract global model features and then outputs possible global symmetry parameters, where input shapes are represented using voxels. We introduce a dedicated symmetry distance loss along with a regularization loss to avoid generating duplicated symmetry planes. Our network can also identify generalized cylinders by predicting their rotation axes. We further provide a method to remove invalid and duplicated planes and axes. We demonstrate that our method is able to produce reliable and accurate results. Our neural network based method is hundreds of times faster than the state-of-the-art methods, which are based on sampling. Our method is also robust even with noisy or incomplete input surfaces.
		\end{abstract}
		
		}

	\maketitle

	\IEEEdisplaynontitleabstractindextext

	%
	\IEEEpeerreviewmaketitle

	\ifCLASSOPTIONcompsoc
	\IEEEraisesectionheading{\section{Introduction}\label{sec:introduction}}
	\else
	\section{Introduction}
	\label{sec:introduction}
	\fi
	
	\IEEEPARstart
	{T}{he} vast majority of species and man-made objects exhibit symmetrical patterns. For example,  bilateral symmetry and radial symmetry are two common patterns existing in starfish and sunflower shapes.
	Symmetry is also an important concept in mathematics. An object is symmetric if some properties do not change under certain transformations. In geometric processing, finding symmetries in geometric data, such as point clouds, polygon meshes and voxels, is an important problem, because numerous applications take advantage of symmetry information to solve their tasks or improve the algorithms, \textit{e.g.}, shape matching~\cite{kazhdan2004symmetry}, segmentation~\cite{podolak2006planar}, completion~\cite{simari2006folding}, \textit{etc.}
	
	Detecting symmetry of 3D objects is an essential step for many applications. Among all the symmetry types, the most common and important one is planar reflective symmetry. In a simple case, a shape can be aligned to the principal axes by applying principal component analysis (PCA). Then planes formed by pairs of principal axes (\textit{i.e.}, orthogonal to the remaining principal axis) can be checked to see if they are symmetry planes. This simple approach works for simpler cases, where the symmetry plane is well aligned with principal axes. However, it is unable to detect symmetry planes which are not orthogonal to any principal axis (\textit{e.g.}, any plane that passes through the rotation axis of a cylinder is a symmetry plane which may not be orthogonal to any principal axis).
	Moreover, the method is highly sensitive to even small changes of geometry, leading to poor detection results. The state-of-the-art methods for symmetry plane detection~\cite{kazhdan2002reflective,korman2014probably} are based on spatial sampling which is much more robust than PCA. However, since they need to sample many potential candidates, the output may produce poor or inaccurate results depending on the random sampling.
	
	To address such limitations, in this paper, we introduce a novel learning framework to automatically discover global planar reflective symmetry. We use a deep convolutional neural network (CNN) to extract the global model feature and capture possible global symmetry. To make CNN-based learning easier and more effective, we convert shapes in arbitrary representations to voxels, before feeding them into our network. The output of our network involves parameters representing reflection planes and rotation axes. Although our aim is to detect planar reflective symmetry, for surfaces of revolution such as a cylinder, any plane passing through the rotation axis can be a symmetry plane. By detecting rotation axes explicitly, we can ensure such symmetry planes are fully detected.
	Our network is unsupervised because we do not require any annotations for symmetry planes of the target objects. This makes collecting training data much easier.
	To achieve this, we introduce a novel symmetry distance loss and a regularization loss to effectively train our network.
	The former measures deviation of the geometry from symmetry given a potential symmetry plane, and the latter
	is used to avoid generating duplicated symmetry planes.
	We further provide a post-processing method to remove invalid and duplicated planes and axes.
	Compared with the method of Kazhdan~\etal ~\cite{kazhdan2002reflective}, our method can produce more reliable and accurate results. More importantly, our learning based approach is hundreds of times faster, achieving real-time performance.
	We also show our network is robust even for noisy and incomplete input.
	Our key contributions are:
	\begin{itemize}
		\item We develop PRS-Net, the first unsupervised deep learning approach to detecting global planar reflective symmetry of 3D objects. Our approach is hundreds of times faster than state-of-the-art methods and also more accurate and reliable.
		\item We model the symmetry detection problem as a differentiable function, which can be attained by a neural network. We further design a dedicated symmetry distance loss along with a regularization loss to avoid generating duplicated symmetry planes. Thanks to the loss functions, our network is trained in an unsupervised manner, making data collection much easier.
	\end{itemize}
	
	\section{Related Work}
	
	\subsection{Symmetry Detection}
	
	Symmetry detection is an important topic in shape analysis, and is widely used in 2D images and 3D geometry. Symmetry detection includes global and partial symmetry, as well as intrinsic and extrinsic symmetry. Most methods can cope with certain level of approximate symmetry.
	
	For 2D image symmetry detection, traditional methods usually vote on the parameters of the symmetry by matched points and angles.
	Marola~\cite{marola1989detection} introduces a metric named coefficient of symmetry for finding symmetry axes of 2D images.
	Sun and Si~\cite{sun1999fast} use Fourier analysis to find symmetry axes of images. Loy and Eklundh~\cite{loy2006detecting} introduce a simple method for grouping symmetric constellations of features. Some methods~\cite{liu2010curved,lee2011curved} detect curved reflection symmetry. Hauagge and Snavely~\cite{hauagge2012image} propose a method to extract multi-scale local bilateral and rotational symmetries. There are also some pioneer research works for 2D symmetry detection based on neural networks. Zielke \etal~\cite{zielke1992intensity} introduce a feedforward network named SEED to find the contours of symmetric objects. Fukushima and Kikuchi~\cite{fukushima2006symmetry} propose a method to find symmetry axes using a neural network. Funk and Liu~\cite{funk2017beyond} introduce the Sym-Net to detect reflection and rotational symmetry. Vasudevan \etal~\cite{vasudevan2018deep} handle symmetry classification using a deep convolutional neural network in Fourier space. These methods focus on 2D image symmetry detection, rather than symmetry on 3D shapes.
	
	For symmetry detection of 3D objects, Atallah~\cite{atallah1985symmetry} proposes an algorithm for enumerating all the axes of symmetry of a planar figure consisting of simple components such as segments, circles and points. Martinet \etal~\cite{martinet2006accurate} propose a method for detecting global accurate symmetry using generalized moments. Kazhdan \etal~\cite{kazhdan2002reflective} detect $n$-fold rotational symmetry based on the correlation of the spherical harmonic coefficients. Raviv \etal~\cite{raviv2010full} present a generalization of symmetries for non-rigid shapes.
	
	Based on whether the symmetry exists in the (Euclidean) embedding space, or based on distance metrics of the geometry, symmetry can be classified as extrinsic and intrinsic.
	
	For extrinsic symmetry, we usually use the Euclidean distance between points to measure the symmetry of a shape, while intrinsic symmetry is measured by different metrics. For global extrinsic symmetry, planar reflective symmetry is the most fundamental one. Zabrodsky \etal~\cite{zabrodsky1995symmetry} introduce a measure of approximate symmetry. Podolak \etal~\cite{podolak2006planar} further describe a planar reflective symmetry transform (PRST) that captures a continuous measure to help find the reflective symmetry. Li \etal~\cite{li2016efficient} present a detection method based on the viewpoint entropy features. Ecins \etal~\cite{ecins2017detecting} introduce a method that mainly detects symmetric objects in 3D scenes and scans of real environment. Cicconet \etal~\cite{cicconet2017finding} regard the problem of finding planar symmetry as a problem of registering two datasets. Ji and Liu~\cite{ji2019fast} propose a network to detect reflection symmetry, but their method needs supervised learning with annotated data. In contrast, our method is unsupervised, without requiring annotated data for training.
	
	For intrinsic symmetry detection, Ovsjanikov \etal~\cite{ovsjanikov2008global} introduce a method to compute intrinsic symmetry of a shape using eigenfunctions of Laplace-Beltrami operators. Kim \etal~\cite{kim2010mobius} present a method to discover point correspondences to detect global intrinsic symmetry on 3D models based on the algorithm by Lipman \etal~\cite{lipman2009mobius}. Mitra \etal~\cite{mitra2010intrinsic} present a method to detect intrinsic regularity, where the repetitions are on the intrinsic grid.
	
	For partial symmetry detection, a shape $S$ is said to have partial symmetry w.r.t. a transformation $T$, if there are two subsets $S_1, S_2\subset S$ such that $T(S_1)=S_2$. Gal and Cohen-Or~\cite{gal2006salient} introduce local surface descriptors that represent the geometry of local regions of the surface to detect partial symmetry. Mitra \etal~\cite{mitra2006partial} present a method based on transformation space voting schemes to detect partial and approximate symmetry. Pauly \etal~\cite{pauly2008discovering} present a method for discovering regular or repeated geometric structures in 3D shapes.
	Berner \etal~\cite{berner2008graph} present a symmetry detection algorithm based on analyzing a graph of surface features. Lipman \etal~\cite{lipman2010symmetry} introduce the Symmetry Factored Embedding (SFE) and the Symmetry Factored Distance (SFD) to analyze and represent symmetries in a point set.
	Xu \etal~\cite{xu2009partial} extend \emph{PRST}~\cite{podolak2006planar} to extract partial intrinsic reflective symmetries.
	
	Our aim is to develop an unsupervised deep learning approach for effective \emph{real-time} global planar reflective symmetry detection.
	\begin{figure*}[t]
		\begin{center}
			\includegraphics[width = \textwidth]{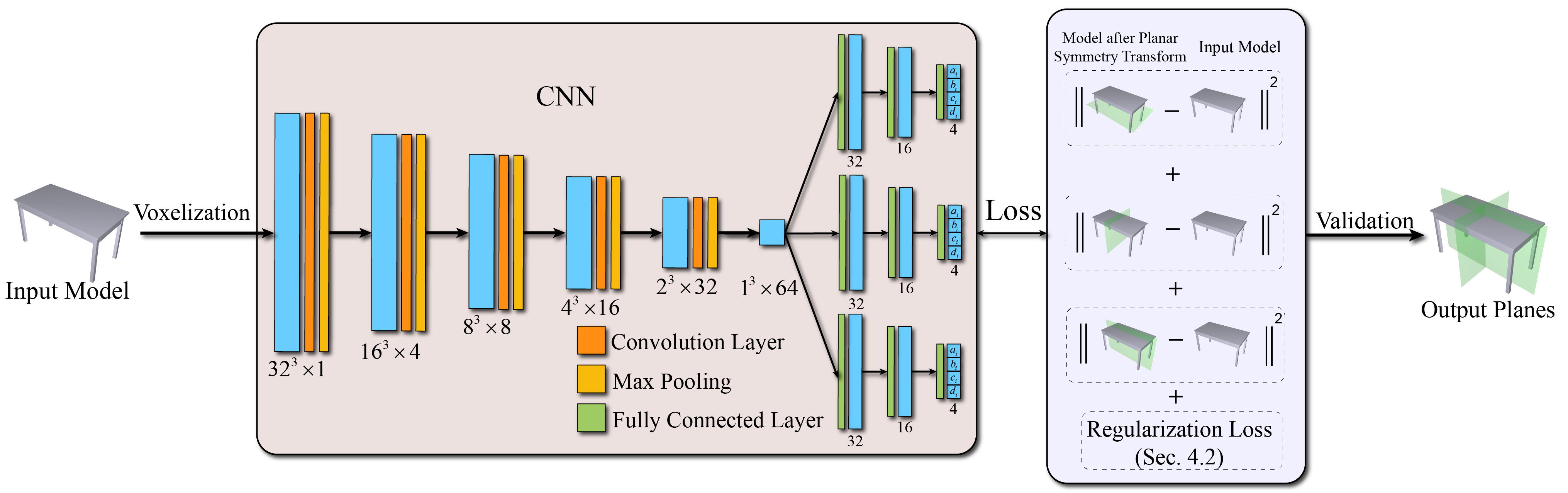}
		\end{center}
		\caption{
			Overview of our method. The input to the network is the voxelized volume of the mesh model. A CNN is used to predict the parameters of the symmetry planes and the planar reflective symmetry transforms associated with the symmetry planes define the loss to train the CNN. This makes the training of CNN operate in an unsupervised manner without any labeled data.  The regularization loss is used to avoid predicting repetitive symmetry planes.
			To simplify the network architecture, our network predicts a fixed number (three in practice) of symmetry planes and rotation axes, which may not all be valid.
			Duplicated or invalid symmetry planes are removed in the validation stage.}
		\vspace{-3mm}
		\label{network}
	\end{figure*}
	\subsection{Geometry Processing with Deep Learning}
	Neural networks have achieved much success in various areas. In recent years, more and more researchers generalize this tool from 2D images to 3D geometry. Su \etal~\cite{su2015multi} use a dimensionality reduction strategy that puts 2D rendered images of a 3D object from multiple views into several classical and mature 2D CNNs. Maturana~\cite{maturana2015voxnet} argues that many existing systems do not take full advantage of 3D depth information, so they create a volumetric occupancy grid representation and predict 3D targets in real time directly from the 3D CNN. Wu \etal~\cite{wu20153d} introduce 3D ShapeNet to learn the distribution of complex 3D voxel grids and use it for 3D shape classification and recognition. Girdhar \etal~\cite{girdhar2016learning} propose a TL-embedding network to generate 3D voxel models from 2D images. Qi \etal~\cite{qi2016volumetric} combine a volumetric CNN with a multi-view CNN, enabling it to be used for object classification of 3D data. Wu \etal~\cite{wu2016learning} propose 3D-GAN (Generative Adversarial Network) to generate 3D objects from a probabilistic space. Tulsiani \etal~\cite{tulsiani2017learning} present a network to interpret 3D objects with a set of volumetric primitives. Riegler \etal~\cite{riegler2017octnet} present OctNet for 3D object classification, using an octree structure to reduce memory and computational costs. Liu \etal~\cite{liu2018planenet} present a network for piece-wise planar reconstruction from a single RGB image.
	
	For point cloud representation, Qi \etal~\cite{qi2017pointnet} introduce PointNet for classification and segmentation. Engelmann \etal~\cite{engelmann2017exploring} build a network based on PointNet to deal with large-scale scene semantic segmentation. Wang \etal~\cite{wang2019dynamic} propose a new module named EdgeConv, which acts on dynamic graphs for CNNs, and encodes local neighborhood information. Wang and Solomon~\cite{wang2019deep} propose a learning-based method named Deep Closest Point (DCP) to predict rigid transformation for point cloud registration, and a Partial Registration Network (PRNet)~\cite{wang2019prnet} for partial-to-partial registration. Meng \etal~\cite{meng2019vv} propose a network which applies group convolutions on regular voxel grids and encodes features computed from radial basis functions (RBF), for point cloud segmentation, achieving state-of-the-art results. Besides, there are some methods that deal with graphs and mesh representations. Hanocka \etal~\cite{hanocka2019meshcnn} propose MeshCNN, which defines mesh convolution and pooling on the mesh edges to handle various tasks including mesh classification and semantic segmentation. Gao \etal~\cite{gao2018automatic} introduce VAE-Cycle-GAN for automatic unpaired shape deformation transfer. Some methods analyze  man-made objects using tree or graph structures to handle model reconstruction, generation and interpolation, such as~\cite{li2017grass,mo2019structurenet,gao2019sdm}.
	
	Such works show great potential for using deep learning for 3D geometry processing, but none of the existing work considers learning to detect 3D object symmetry in an unsupervised manner, which we will address in this paper.

	\section{Network Architecture}
	In this section, we describe the network architecture of our method. 	The overall network is presented in Figure~\ref{network}. This work aims to train a CNN to predict the symmetry planes in an unsupervised manner. The CNN has five 3D convolution layers of kernel size 3, padding 1, and stride 1. After each 3D convolution,  a max pooling of kernel size 2 and leaky ReLU~\cite{maas2013rectifier} activation are applied. These are followed by fully connected layers to predict the parameters of symmetry planes.
	
	The input of the network is $32\times 32\times 32$ voxels which are voxelized from the input shape.
	As we will later show in Section~\ref{sec:res}, this resolution achieves better performance than alternative settings.
	
	The 3D convolution and pooling are used to extract global features of the shape. The output includes parameters of reflective planes and rotation parameters. For typical shapes, our network predicts three potential symmetry planes and three potential rotation axes. These will be further validated in the validation stage so the shape may have fewer (or even none) symmetry planes. The symmetry planes $\mathbf{P}_i$ ($i=1, 2, 3$) are represented using an implicit representation. We further use quaternions to represent rotations $\mathbf{R}_i$ ($i=1, 2, 3$), because the quaternion is more compact with fewer parameters compared to the rotation matrix, and it can be easily transformed from and to an axis-angle representation.
	We initialize the normal vectors of the planes and the directions of the axes to be three vectors perpendicular to each other to maximize their coverage.
	In practice, we simply set $\mathbf{v_1}=(1,0,0),\mathbf{v_2}=(0,1,0),\mathbf{v_3}=(0,0,1)$ to initialize them. The initial angle of each rotation axis is set to $\theta=\pi$, thus the corresponding quaternion is $\mathbf{R}_i=\cos(\theta/2)+\sin(\theta/2)(\mathbf{v}_{i}^{(1)}\mathbf{i}+\mathbf{v}_i^{(2)}\mathbf{j}+\mathbf{v}_i^{(3)}\mathbf{k})$, where $\mathbf{i}$, $\mathbf{j}$, and $\mathbf{k}$ are the fundamental quaternion units, $\theta$ is the rotation angle, and $\mathbf{v}_i^{(k)}$ is the $k^{\rm th}$ component of $\mathbf{v}_i$ ($k=1, 2, 3$ corresponding to $x$, $y$ and $z$). In our network, the predicted quaternions $\mathbf{R_i}$ are normalized to a unit vector after each iteration of optimization.
	After the training is finished, we transform the quaternion representation to the axis-angle representation.
	
	Our network is trained in an unsupervised manner because we do not require any annotations for the reflection plane parameters that best describe the global extrinsic symmetry of the object. This greatly reduces the effort of obtaining training data as only a collection of (symmetric) shapes is required. In order to achieve this, we propose a novel symmetry distance loss to promote planar symmetry. Moreover, to avoid producing duplicated symmetry planes, we further introduce a regularization loss.

	\section{Loss Function}
	Denote by $\mathbf{P}_i=(\mathbf{n}_i,d_i)$ the $i^{\rm th}$ plane parameters in implicit form, where $\mathbf{n}_i=(a_i,b_i,c_i)$ is the normal direction of the plane, which uniquely determines a reflection plane $a_{i}x+b_{i}y+c_{i}z+d_i=0$, and $\mathbf{R}_i=(u_{i0},u_{i1},u_{i2},u_{i3})$ is the $i^{\rm th}$ rotation parameters, which represent the quaternion $u_{i0}+u_{i1}\mathbf{i}+u_{i2}\mathbf{j}+u_{i3}\mathbf{k}$ of rotation transform. To train the network to predict symmetry planes and rotation axes, we introduce two loss functions, namely symmetry distance loss and regularization loss.
	
	\subsection{Symmetry Distance Loss}\label{sdl} To measure whether an input shape $O$ is symmetric w.r.t. a given reflection plane or a rotation axis, we first uniformly sample $N$ points on the shape to form a point set $\mathcal{Q}$. We then obtain a transformed  point set $\mathcal{Q}{'}$ by applying planar symmetry or rotation transformation to each point $\mathbf{q}_k \in  \mathcal{Q}$ to obtain the transformed point $\mathbf{q}_k'$.
	
	For the $i^{\rm th}$ reflection plane, the symmetry point $\mathbf{q}'_k$ of point $\mathbf{q}_k$ is:
	\begin{equation}
	\mathbf{q}'_k=\mathbf{q}_k-2\frac{\mathbf{q}_k\cdot \mathbf{n}_i+d_i}{\|\mathbf{n}_i\|^2}\mathbf{n}_i,
	\end{equation}
	where $\mathbf{n}_i$ is the normal vector of the $i^{\rm th}$ reflection plane, and $d_i$ is its offset parameter.
	
	For the $j^{\rm th}$ rotation axis, the symmetry point $\mathbf{q}'_k$ of point $\mathbf{q}_k$ is
	\begin{equation}
	\hat{\mathbf{q}}'_k=\mathbf{p}_j\hat{\mathbf{q}}_k\mathbf{p}_j^{-1},
	\end{equation}
	where $\mathbf{p}_j=(p_{j0},p_{j1},p_{j2},p_{j3})$ represents the quaternion of the $j^{\rm th}$ rotation, and $\hat{\mathbf{q}}_k=(0,\mathbf{q}_k)$ is the quaternion form of point $\mathbf{q}_k$. This results in a new quaternion $\hat{\mathbf{q}}'_k$, with ${\mathbf{q}}'_k$ as its imaginary part.

	Then we calculate the shortest distance
	\begin{equation}\label{eq:dk}
	D_k=\min_{\mathbf{o}\in O}\|\mathbf{q}'_k-\mathbf{o}\|_2
	\end{equation}
	from symmetry points $\mathbf{q}'_k$ to the shape $O$. To calculate $D_k$ efficiently, we precompute the closest point on the surface to each grid center point of a regular grid, and during training, we calculate the distance between symmetry points to the corresponding closest point in the same grid as the approximate closest distance and their gradients required for back propagation.
	Finally, the symmetry distance loss of a shape is defined as
	\begin{equation}
	L_{sd}=\sum_{i=1}^{3}\sum_{k=1}^{N} \hat{D}_k^{(i)} + \sum_{j=1}^{3}\sum_{k=1}^{N} \tilde{D}_k^{(j)}
	\end{equation}
	where $N=|\mathcal{Q}|$ is the number of sample points, $\hat{D}_k^{(i)}$ and $\tilde{D}_k^{(j)}$ represent the symmetry error for the $k^{\rm th}$ sample point w.r.t. the $i^{\rm th}$ symmetry plane and the $j^{\rm th}$ rotation axis, respectively, as defined in Eq.~\ref{eq:dk}.
	\begin{figure}[t]
		\begin{center}

			\includegraphics[width = 0.7\linewidth]{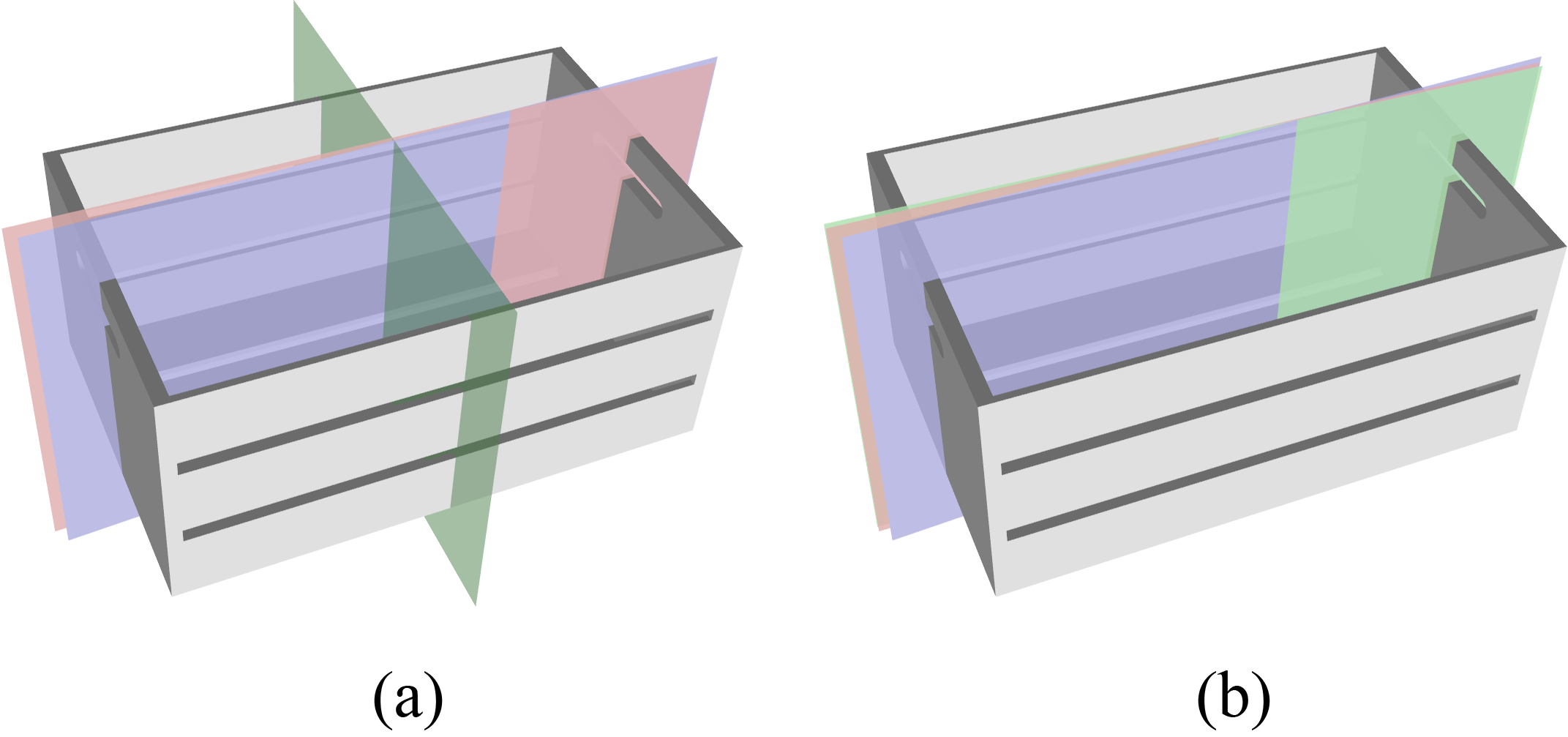}
		\end{center}\vspace{-4mm}
		\caption{
			The results of the network with (a) and without (b) the regularization loss. We train the network with the same initial plane parameters.
			Without the regularization loss, the network cannot differentiate near identical planes, as they also achieve a local minimum of the loss function during training. Our regularization loss effectively promotes non-identical outputs, which helps to cover different symmetry planes.}
		\vspace{-4mm}
		\label{constrain}
	\end{figure}
	\subsection{Regularization Loss} Many shapes have multiple symmetry planes or rotation axes. Since each of them is sufficient to minimize the symmetry distance loss, the network may produce multiple, near-identical outputs (\textit{e.g.} $\mathbf{P}_1 \approx \mathbf{P}_2$). This may lead to an output that misses essential symmetry planes/rotation axes.
	To address this, we aim to constrain the learning of reflection planes and rotation axes to not overlap with each other, by adding a regularization loss $L_{r}$ to separate each plane and axis from each other as much as possible.
	Let $\mathbf{M}_1$ be a $3\times 3$ matrix where each row is the unit normal direction of a symmetry plane, \textit{i.e.}, $\mathbf{M}_1 = [\frac{\mathbf{n}_1}{\|\mathbf{n}_1\|} \quad \frac{\mathbf{n}_2}{\|\mathbf{n}_2\|} \quad \frac{\mathbf{n}_3}{\|\mathbf{n}_3\|}]^T$. Let $\mathbf{M}_2$ be another $3\times 3$ matrix where each row contains the normalized axis direction  of rotational symmetry prediction. Let
	\begin{eqnarray}
	&\mathbf{A} = \mathbf{M}_1 \mathbf{M}_{1}^T - \mathbf{I},\\
	&\mathbf{B} = \mathbf{M}_2 \mathbf{M}_2^T - \mathbf{I},
	\end{eqnarray}
	where $\mathbf{I}$ is the $3\times 3$ identity matrix. If each plane (resp. axis) is orthogonal to every other plane (resp. axis), then $\mathbf{A}$ and $\mathbf{B}$ are all-zero matrices. We define the regularization loss as
	\begin{equation}
	L_{r}=\|\mathbf{A}\|_F^2 + \|\mathbf{B}\|_F^2 = \sum_{i=1}^{3}\sum_{j=1}^{3}(\mathbf{A}_{ij}^2+\mathbf{B}_{ij}^2),
	\end{equation}
	which penalizes planes and axes closer to parallel, where $\|\cdot\|_F$ is the Frobenius norm.
	Figure~\ref{constrain} compares the results with (a) and without (b) the regularization loss. This regularization term is the soft constraint and would not enforce the planes to be strictly perpendicular, as shown in Figure~\ref{fig:para}.
	In this experiment, we initialize all the planes and axes with the same settings. As can be seen,  the reflection planes overlap with each other without the regularization loss as shown in the right column, because it is difficult for the network to separate them as they also achieve the same local minimum, while the planes are clearly separated on the left thanks to the regularization. There are still two overlapping planes because the model does not have the third reflection plane with small symmetry distance loss (which will be addressed in validation).
	
	\subsection{Overall Loss Function}
	
	We define the overall loss function as
	\begin{equation}
	L = L_{sd} + w_r L_r,
	\end{equation}
	where $w_r$ is a weight to balance the importance of two loss terms.
	
	\begin{figure*}[t]
		\begin{center}
			\includegraphics[width = \linewidth]{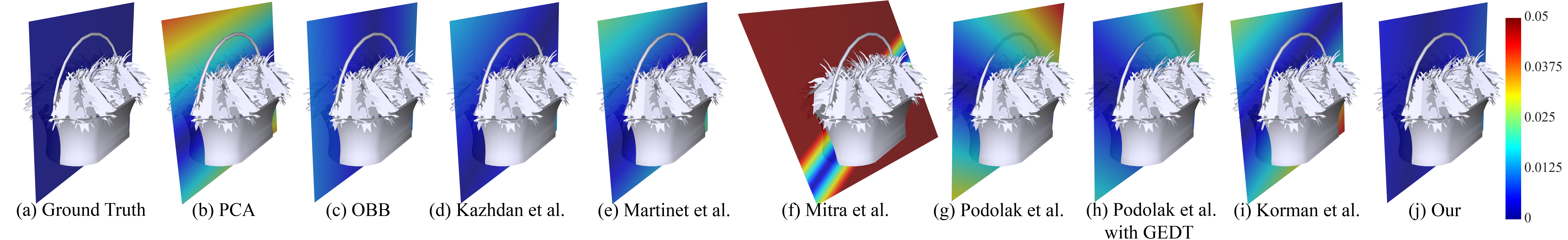}
		\end{center}\vspace{-4mm}
		\caption{Comparison of symmetry plane detection using different methods. The heatmap describes the distance between the ground truth and the plane of each correspond method. It shows that our method has less error and more reasonable results than other methods.}
		\label{compare2}\vspace{-4mm}
	\end{figure*}
	\begin{figure}[h]
		\begin{center}
			\includegraphics[width = 0.8\linewidth]{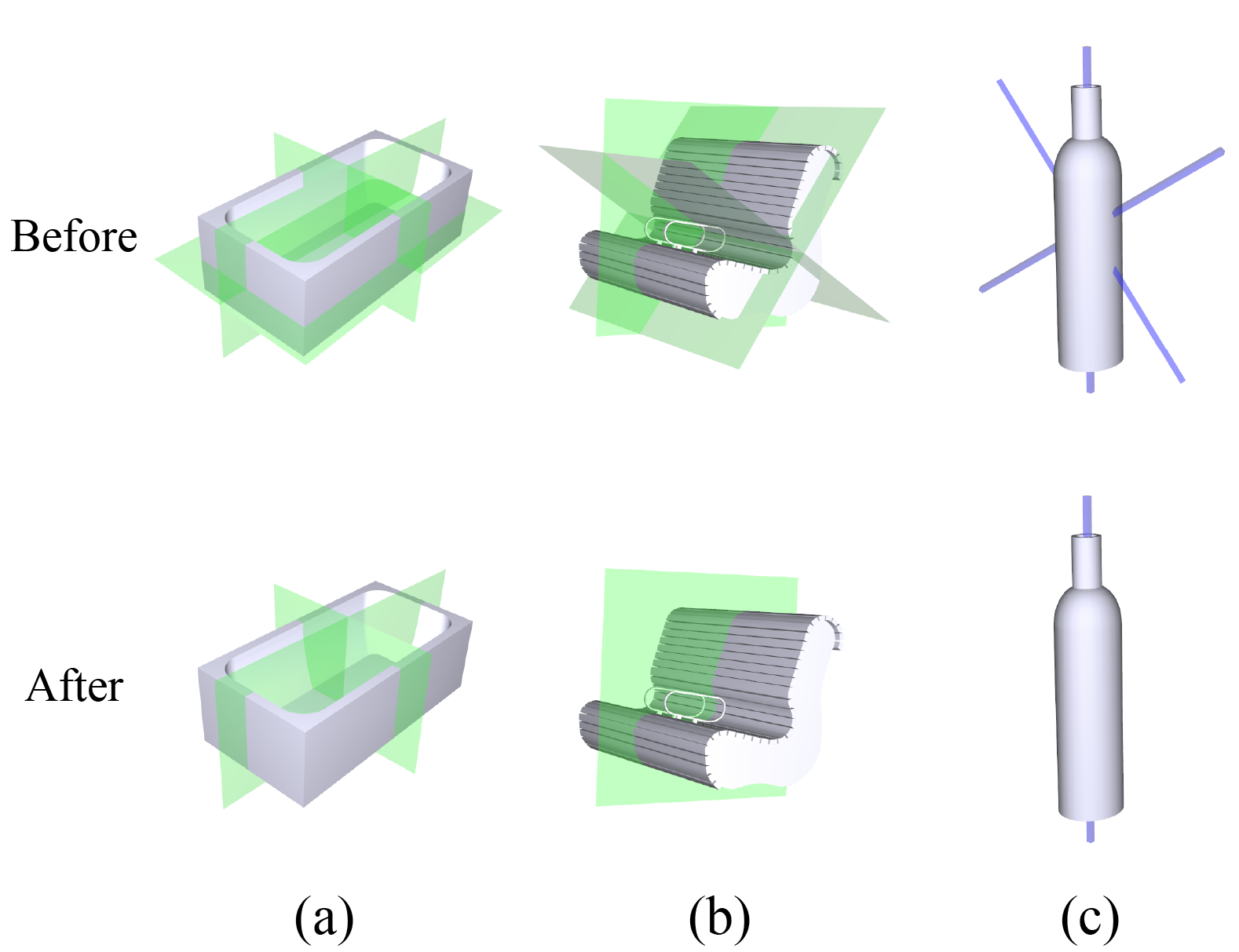}
		\end{center}\vspace{-4mm}
		\caption{The results before (the first row) and after (the second row) validation. Most models have fewer than 3 symmetry planes/rotation axes, and the initial network output may contain extra symmetry outputs (which are local minima or duplicates). Our simple validation stage removes duplicated outputs as well as symmetry outputs which are not of acceptable quality.}
		
		\label{postprocess}\vspace{-4mm}
		
	\end{figure}

	\section{Validation}
	Our network predicts symmetry planes/rotation axes and it always predicts three symmetry planes and three rotations to simplify the architecture. However, real-world shapes may have fewer symmetry planes and rotation axes. In this case, some output planes may overlap with each other. Moreover, due to the local minimum nature of gradient descent optimization, the network may also detect some approximate symmetry which is not sufficiently good. These issues however can be easily addressed by a simple validation stage. {We check the detected symmetry planes and rotation axes to remove duplicated outputs: if its dihedral angle is less than $\pi/6$, we remove the one with larger symmetry distance error. Meanwhile, if the detected symmetry planes/rotation axes lead to high symmetry distance loss (greater than $4\times10^{-4}$ in our experiments), we also remove them as they are not sufficiently symmetric. In particular for rotational symmetry, as we are only concerned with detecting surfaces of revolution with infinite symmetry planes going through the symmetry axis, for each detected rotation axis, we consider rotations by an arbitrary angle (every $1^\circ$ in practice), and only accept it if the rotational symmetry is retained with arbitrary rotation angles.
		
		Note that we normalize the shapes to fit in a unit cube, so this threshold is generally applicable, and obtained by working out the distributions of symmetry distance errors for a small number of valid cases.
		By normalizing the shapes to the unit cube, this threshold is applicable for all the shapes.
		As shown in Figure~\ref{postprocess}(a), the bath only has two reflection planes, but the network always outputs three symmetry planes before validation. Their symmetry distance errors are $6.67\times10^{-5}, 9.43\times10^{-5}$ and $2.57\times10^{-3}$ respectively, so the validation removes the third plane and retains the other two planes. The symmetry distance errors of the three output planes of the bench in Figure~\ref{postprocess}(b) are $9.50\times10^{-6}$, $1.46\times10^{-3}$ and $1.31\times10^{-3}$, so two planes are removed due to high symmetry distance loss. Similarly, two extra rotation axes are removed in Figure~\ref{postprocess}(c).}

	\begin{figure*}[t]
		\centering
		\subfigure[bench]{
			\label{fig:bench}
			\centering
			\includegraphics[width = 0.9\textwidth]{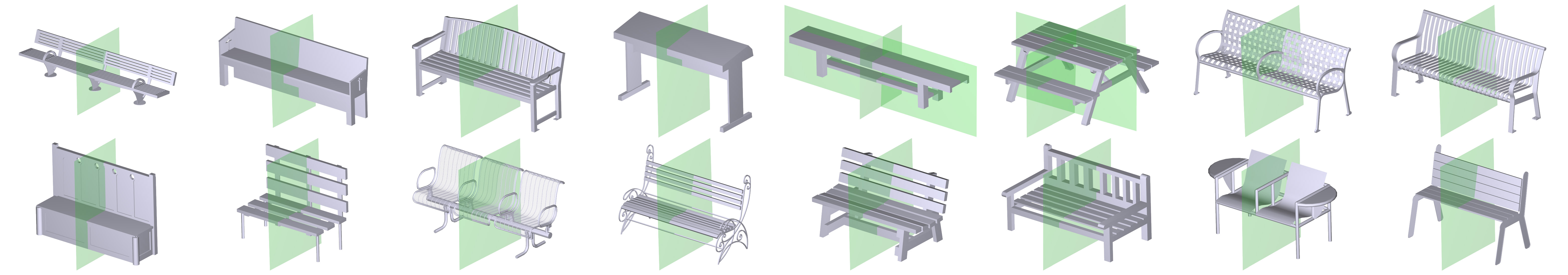}
		}
		\subfigure[plane]{
			\label{fig:plane}
			\centering
			\includegraphics[width = 0.9\textwidth]{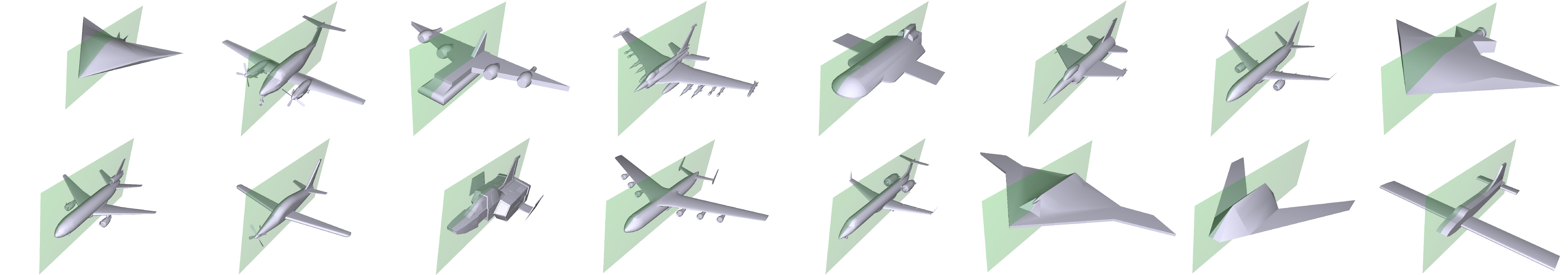}
		}
		
		\subfigure[table]{
			\label{fig:table}
			\centering
			\includegraphics[width = 0.9\textwidth]{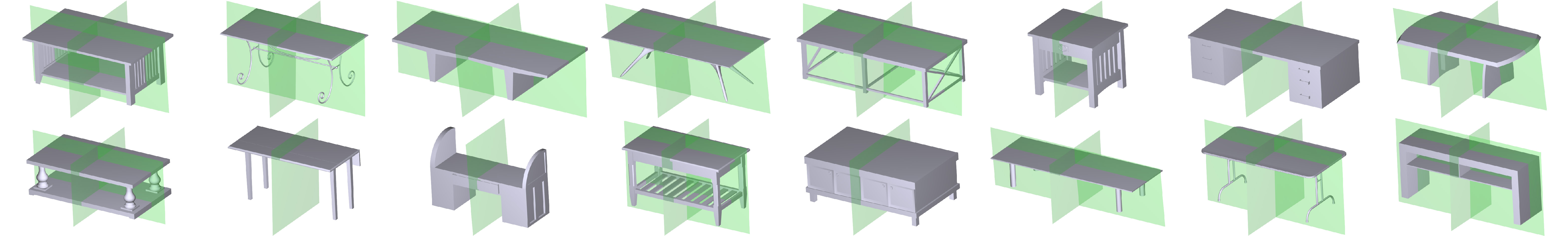}
		}
		\subfigure[boat]{
			\label{fig:boat}
			\centering
			\includegraphics[width = 0.9\textwidth]{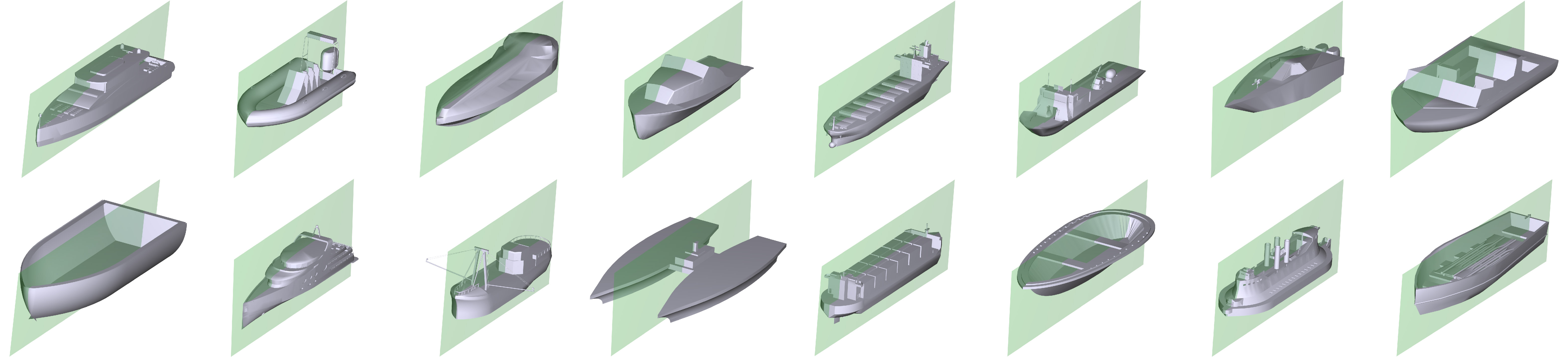}
		}
		\caption{More results of symmetry planes discovered by our method of different shapes from test data in ShapeNet \cite{chang2015shapenet}, including \subref{fig:bench} bench, \subref{fig:plane} plane, \subref{fig:table} table, and \subref{fig:boat} boat.}
		\label{all2}
	\end{figure*}
	\begin{figure}[t]
		\centering
		\includegraphics[width=0.9\linewidth]{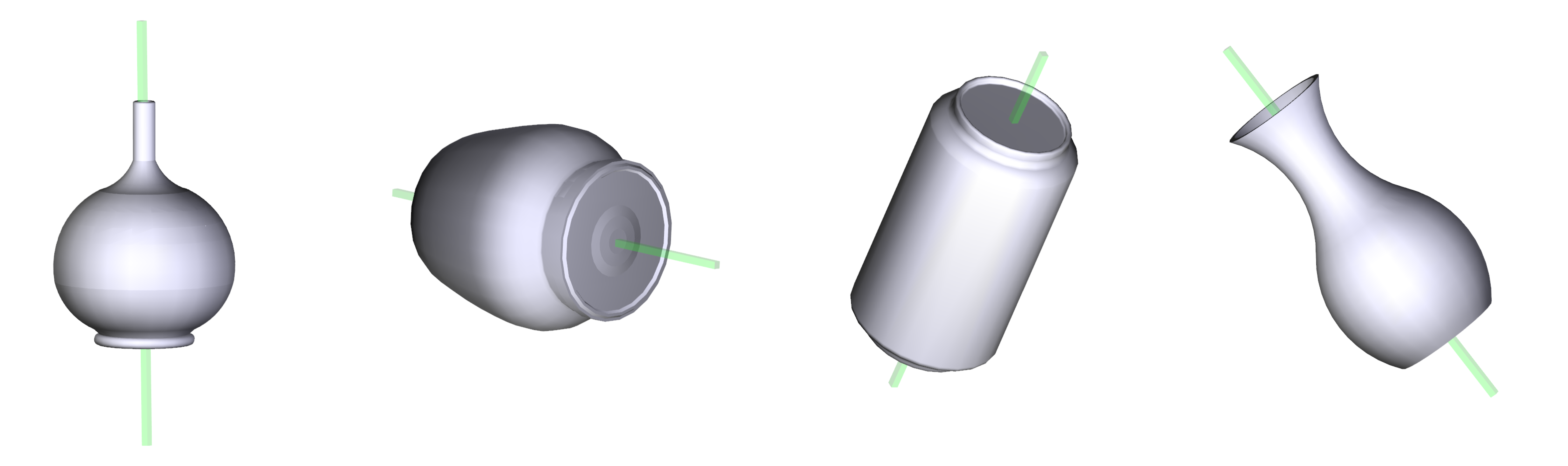}
		\caption{The rotation axes of generalized cylinders detected by our method. Our method can identify these shapes through outputting quaternions that represent rotational symmetry.}\vspace{-4mm}
		\label{iso}
	\end{figure}
	
	\section{Experiments}
	
	\subsection{Dataset and Training}\label{sec:train}
	We use the ShapeNet~\cite{chang2015shapenet} dataset which contains 55 common object categories with about $51,300$ unique 3D models to train our network. We choose $80\%$ models as the training set and the remaining $20\%$ models for testing. Because the ShapeNet models are often roughly axis aligned, and many categories have fewer than $500$ models, we apply different random rotations on each model to obtain $4,000$ augmented models for each category for training.
	However, many shapes in ShapeNet are not symmetric.
	For non-symmetric shapes, it can be difficult to determine which result is better even by human annotators. Although  \textit{e.g.} symmetry distance error (SDE) that measures how close the shape is to itself after mirroring
	could be used to give an indication, it does not provide a definite measure. We are not aware of benchmark datasets for symmetry, and thus we collect a test set containing shapes which are manually validated to ensure they are symmetric, and for these shapes, we obtain reasonable ground truth symmetry planes to compare the errors of different methods w.r.t. the ground truth.
	Because the models in ShapeNet are pre-aligned, we select three axis-aligned planes to be candidates and manually check and correct the results to get the valid planes for each model in the test set as the ground truth. For models with more than three symmetry planes, we also add these planes as the ground truth.
	
	In the preprocessing step, we uniformly sample $1,000$ points on the surface, and generate voxels of size $32\times 32\times 32$ as input to our network. During network training, we set batch size $b=32$, learning rate $l_r=0.01$ and regularization loss weight {$w_{r}=25$}, and then use the ADAM~\cite{kingma2014adam} optimizer to train our network according to the loss described above.

	\subsection{Results and Evaluations}
	We compare our method with PCA, Oriented Bounding Box \cite{chang2011fast}, the methods of Kazhdan \textit{et al.}~\cite{kazhdan2002reflective}, Martinet \textit{et al.}~\cite{martinet2006accurate}, Mitra \textit{et al.}~\cite{mitra2006partial}, Podolak \textit{et al.}~\cite{podolak2006planar}, Podolak \textit{et al.}~\cite{podolak2006planar} with Gaussian Euclidean Distance Transform (GEDT), using their default parameters, and Korman \textit{et al.}~\cite{korman2014probably}.
	To quantitatively evaluate the quality of detected symmetry planes, we first normalize the normal vector of the plane, and adjust the direction of the normal vector to make the angle between the detected normal and ground truth normal no larger than $\pi/2$, and the error of the plane w.r.t. the ground truth (GTE) is defined as
	\begin{equation}
	GTE=(a_i-a_{gt})^2+(b_i-b_{gt})^2+(c_i-c_{gt})^2+(d_i-d_{gt})^2.
	\end{equation}
	Alternatively, we can also use symmetry distance error (SDE) defined in the same way as $L_{sd}$ to measure the symmetry quality.
	
	We compare our method with existing methods. Figure~\ref{compare2} shows the results of reflection planes from different methods. In each figure, the heatmap presents the distance between the sample points on the ground truth plane and the reflection points on each result plane.
	The methods of Kazhdan \textit{et al.}\cite{kazhdan2002reflective} and PRST~\cite{podolak2006planar} are sensitive to the resolution of grids and the distribution of sample points. For PRST, it may also produce different results when it runs multiple times due to the random sampling. The computational time of the method grows quickly when more sample points and higher resolution of grids are used. The last column shows that our method obtains the most accurate result. Moreover, our method only takes a few milliseconds, once the network is trained, while methods except PCA need several iterations to compute the local optimal result. More results as shown in Figure \ref{all2}, our method is able to produce reliable and accurate results, including shapes with multiple symmetry planes. In Figure~\ref{iso}, we show the rotation axes of the generalized cylinders detected by our method. To detect these shapes, we check the symmetry distance error of  rotation axes.
	
	\begin{table*}[t]
		\footnotesize
		\caption{
			The mean ground truth error (GTE) and the symmetry distance error (SDE) of different methods. The reported results are averaged over a validated subset of ShapeNet with $1000$ models from different categories. Our method produces minimum average errors for both SDE and GTE.
		}\vspace{-7mm}
		\begin{center}
			\begin{tabular}{lcccccccccc}
				\toprule
				\multirow{2}{*}{Error} & \multirow{2}{*}{PCA} & {Oriented} & Kazhdan \textit{et.al.} &Martinet \textit{et.al.} &Mitra \textit{et.al.} &\multirow{2}{*}{PRST\cite{podolak2006planar}} & PRST\cite{podolak2006planar} & Korman \textit{et.al.}& \multirow{2}{*}{Our}\\
				&&Bounding Box~\cite{chang2011fast}&\cite{kazhdan2002reflective}&\cite{martinet2006accurate}&\cite{mitra2006partial}&&with GEDT&\cite{korman2014probably}&\\
				\midrule
				GTE($\times 10^{-2}$) & $2.41$ & $1.24$ & $0.17$& $13.6$&$52.1$&$4.42$& $3.97$& $19.2$  & $\mathbf{0.11}$\\
				\midrule
				SDE($\times 10^{-4}$) & $3.32$ & $1.25$ & $0.897$ & $3.95$& $14.2$ &$1.78$& $1.60$&$1.75$& $\mathbf{0.861}$\\
				\bottomrule
			\end{tabular}\vspace{-4mm}
			
		\end{center}
		
		\label{quantity2}
	\end{table*}
	\begin{table*}[t]
		\footnotesize
		\centering
		\caption{
			The symmetry distance error ($\times 10^{-4}$) of different methods on ABC~\cite{Koch_2019_CVPR} and Thingi10K~\cite{zhou2016thingi10k}. Our method produces minimum average SDE and can deal with more complex and asymmetric shapes.
		}\vspace{-7mm}
		\begin{center}
			\begin{tabular}{lcccccccccc}
				\toprule
				\multirow{2}{*}{Dataset} & \multirow{2}{*}{PCA} & {Oriented} & Kazhdan \textit{et.al.} &Martinet \textit{et.al.} &Mitra \textit{et.al.} &\multirow{2}{*}{PRST\cite{podolak2006planar}} & PRST\cite{podolak2006planar} & Korman \textit{et.al.}& \multirow{2}{*}{Our}\\
				&&Bounding Box~\cite{chang2011fast}&\cite{kazhdan2002reflective}&\cite{martinet2006accurate}&\cite{mitra2006partial}&&with GEDT&\cite{korman2014probably}&\\
				\midrule
				ABC~\cite{Koch_2019_CVPR} &6.97  &4.98 &5.54&7.48&12.1&6.28&4.87 &4.54 &$\mathbf{1.14}$\\
				\midrule
				Thingi10K~\cite{zhou2016thingi10k} &7.24 &2.78  &4.34  &4.37& 12.0  &5.43&4.97 &2.11& $\mathbf{1.69}$\\
				\bottomrule
			\end{tabular}\vspace{-4mm}
			
		\end{center}
		
		\label{experiment}
	\end{table*}
	\begin{figure}[t]
		\begin{center}
			\includegraphics[width = \linewidth]{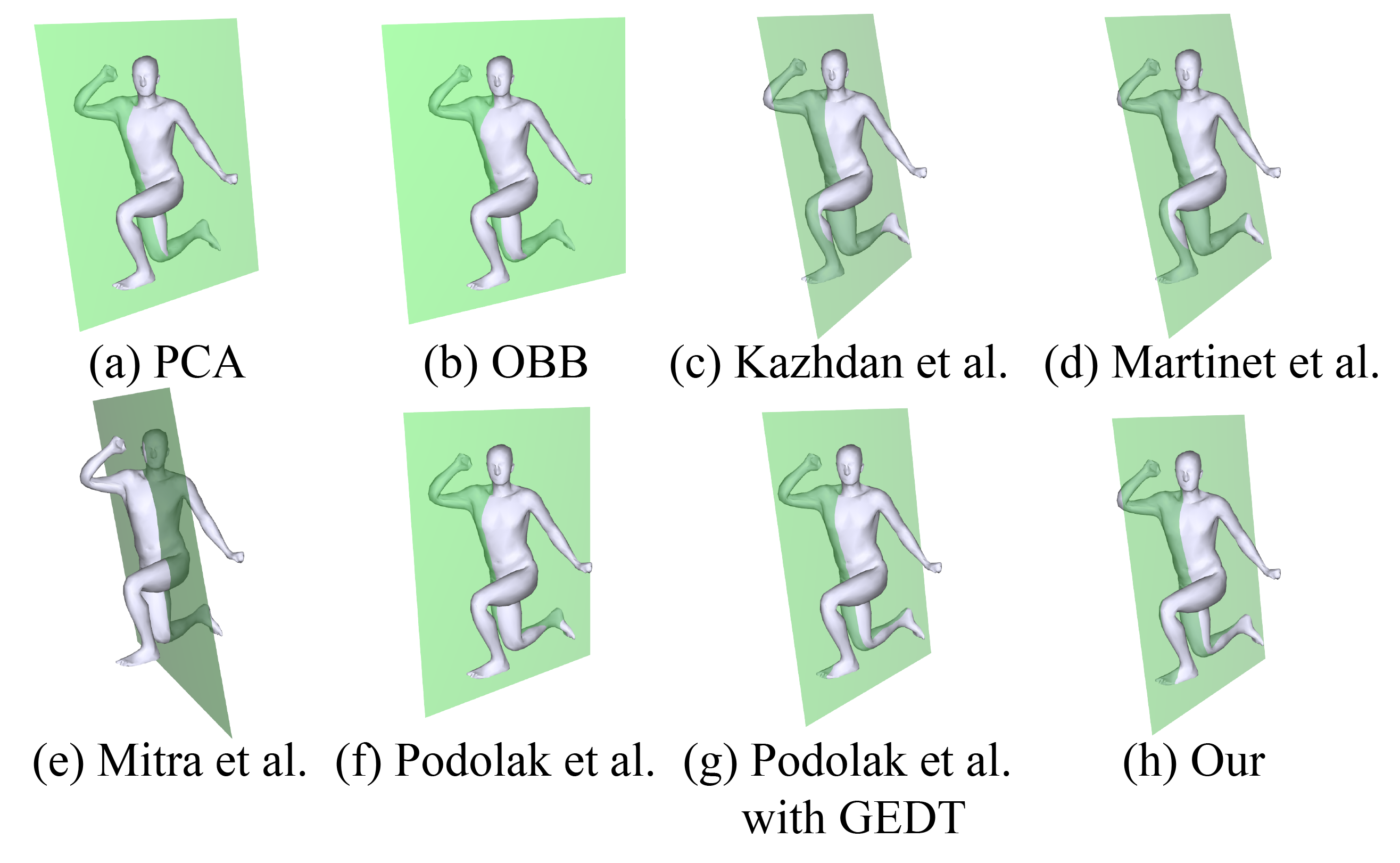}
		\end{center}\vspace{-4mm}
		
		\caption{Comparison of symmetry detection for a human shape from the SCAPE dataset~\cite{anguelov2005scape} only with approximate symmetry using different methods. Our result has the lowest symmetry distance error.}
		\label{scape}\vspace{-4mm}
	\end{figure}
	\begin{figure}[t]
		\centering
		\subfigure[$w_r=1$]{
			\label{fig:w1}
			\includegraphics[width = 0.1\textwidth]{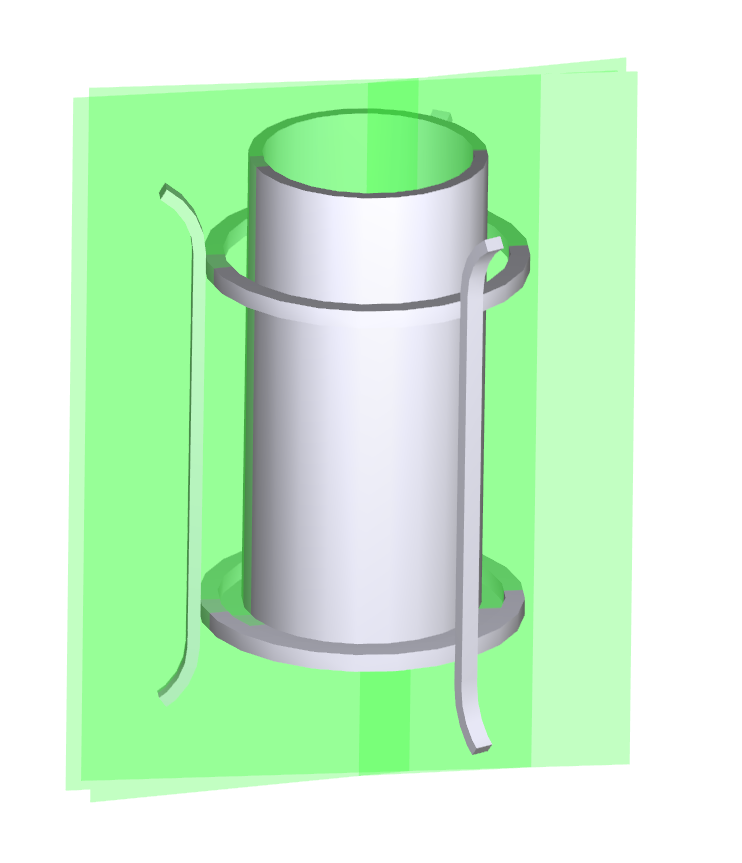}
		}
		\subfigure[$w_r=10$]{
			\label{fig:w10}
			\includegraphics[width = 0.1\textwidth]{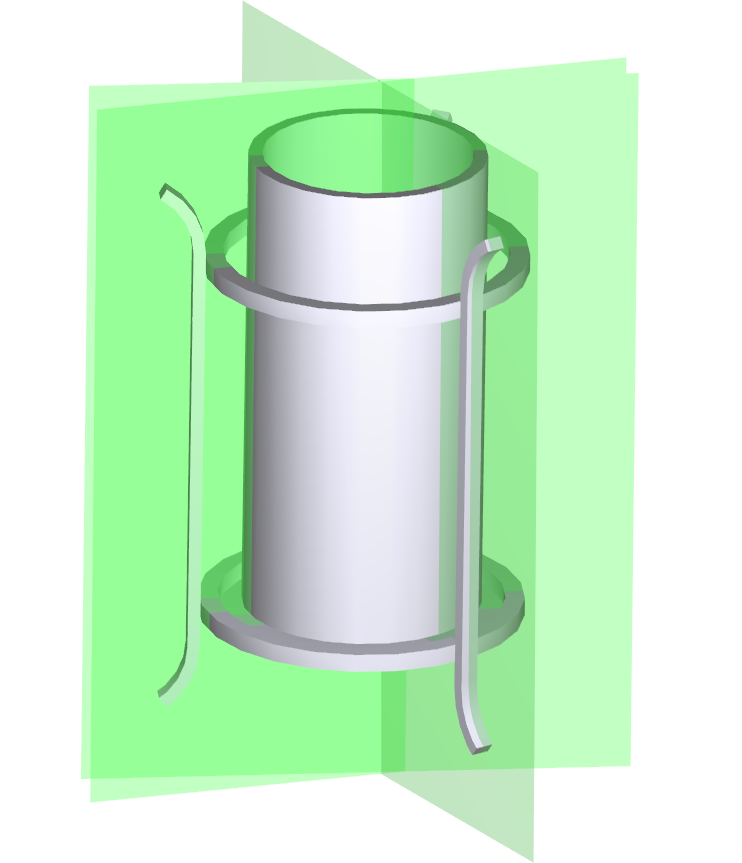}
		}
		\subfigure[$w_r=25$]{
			\label{fig:w25}
			\includegraphics[width = 0.1\textwidth]{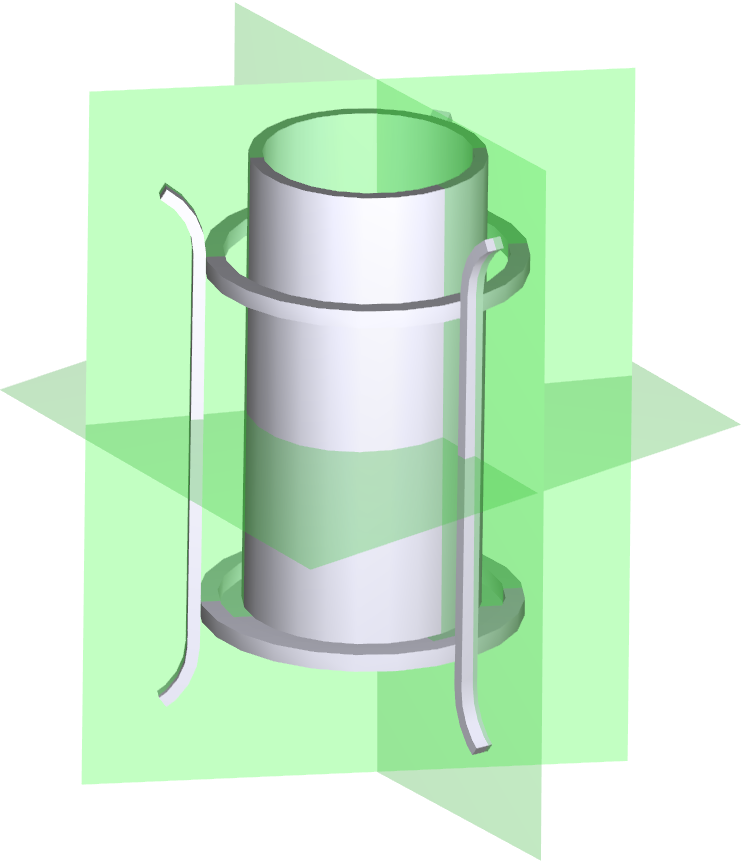}
		}
		\subfigure[$w_r=50$]{
			\label{fig:w50}
			\includegraphics[width = 0.1\textwidth]{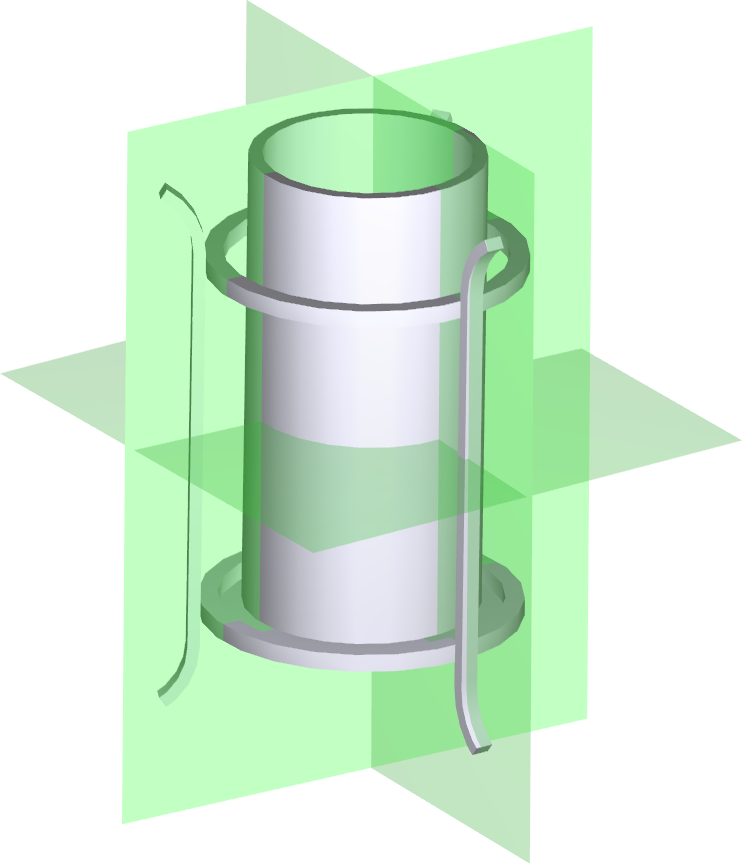}
		}
		\caption{An example of visual results with different regularization weights $w_r$. The table shape has two symmetry planes, and planes overlap with each other when the regularization loss weight $w_r$ is $1$ or $10$, as shown in \subref{fig:w1} and \subref{fig:w10}. The network produces near-identical outputs because the regularization loss with such small weights has little impact and the algorithm identifies duplicated planes so fails to identify the other symmetry plane.  In \subref{fig:w25} and \subref{fig:w50}, the planes are separated with large regularization weights, but \subref{fig:w25} produces more accurate symmetry planes than \subref{fig:w50}.}
		\label{fig:para}\vspace{-4mm}
	\end{figure}
	
	\begin{figure*}[t]
		\begin{center}
			\includegraphics[width = 0.8\linewidth]{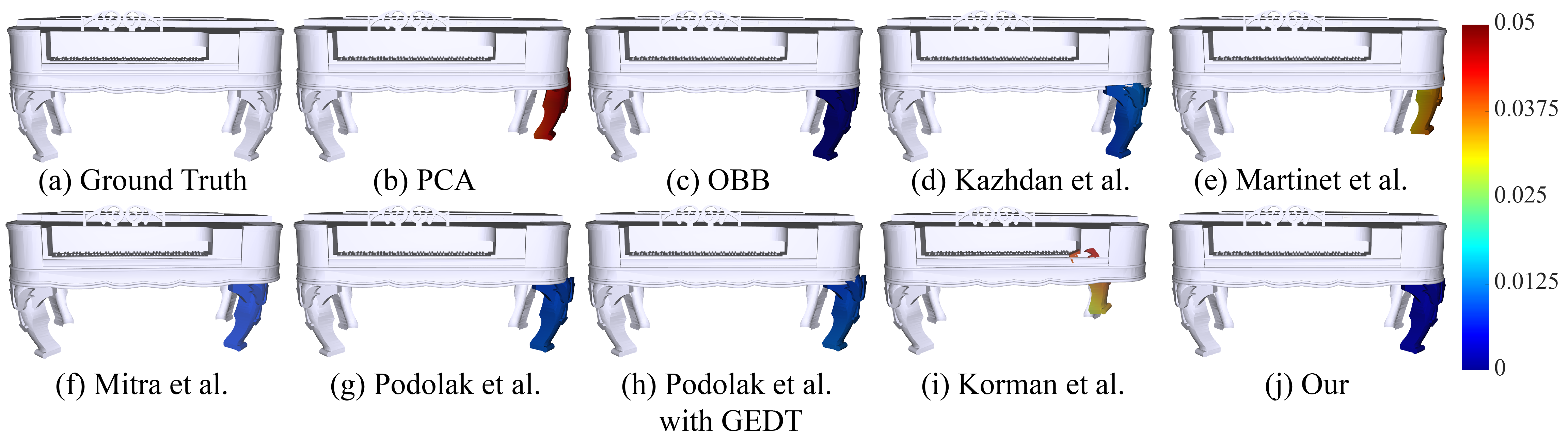}
		\end{center}\vspace{-4mm}
		\caption{Comparison of shape completion.
			Our method repairs the incomplete shape perfectly due to the robustness and accuracy of our method. It shows the Euclidean error between the ground truth part and the generated part of the piano, which is obtained by mirroring the geometry of the left leg along the symmetry planes detected by different methods.}
		\label{completion}
	\end{figure*}

	\begin{figure*}[t]
		\begin{center}
			\includegraphics[width = 0.8\linewidth]{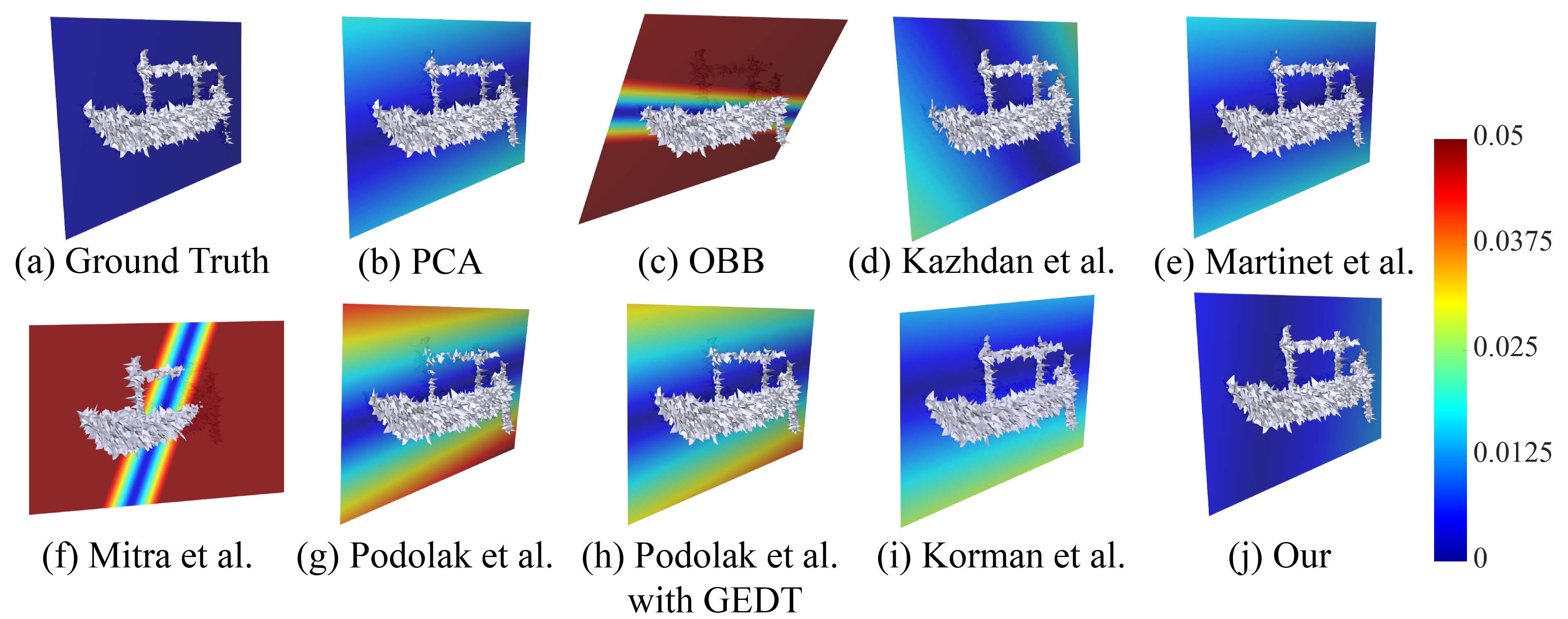}
		\end{center}\vspace{-4mm}
		\caption{Robustness testing with adding Gaussian noise to each vertex along the normal direction. It shows that our method produces stable output.}\vspace{-4mm}
		\label{noise}
	\end{figure*}

	We also evaluate our network on large test sets.
	Table \ref{quantity2} shows the mean ground truth error and symmetry distance error of our test set with $1000$ models, and our method produces minimum error.

	Many shapes are not entirely symmetric, so we evaluate our method on human shapes from the SCAPE dataset~\cite{anguelov2005scape} to test the capability of handling such general cases. As shown in Figure~\ref{scape}, we visualize the symmetry plane with lowest symmetry distance error of each method except for the method of Korman \textit{et al.}~\cite{korman2014probably} which fails to detect any reflective symmetry, and our method discovers a more plausible symmetry plane, which has the lowest symmetry distance error $1.75\times10^{-3}$, compared with $3.49\times10^{-3}$ of PCA, $3.98\times10^{-3}$ of Oriented Bounding Box, $1.83\times10^{-3}$ of the method of Kazhdan \textit{et al.} \cite{kazhdan2002reflective}, $1.90\times10^{-3}$ of the method of Martinet \textit{et al.} \cite{martinet2006accurate}, $4.87\times10^{-3}$ of the method of Mitra \textit{et al.}~\cite{mitra2006partial}, $3.78\times10^{-3}$ of PRST~\cite{podolak2006planar} and $2.01\times10^{-3}$ of PRST with GEDT.
	It demonstrates that our method has the ability to detect \emph{approximate} symmetry planes of general shapes, outperforming existing methods. This is because we use the unsupervised loss to train the network, and the ShapeNet dataset~\cite{chang2015shapenet} has various categories including symmetric and asymmetric shapes.
	Note that our training set does not include any shape from SCAPE or even any human shape. This also shows that our network generalizes well to new shapes and unseen shape categories.

	In Table \ref{experiment}, we also report accuracy comparison of our method with alternative methods (in terms of SDE as no ground truth is available) on ABC~\cite{Koch_2019_CVPR} and Thingi10K~\cite{zhou2016thingi10k} datasets, which contain a large number of asymmetric shapes. In this experiment, we randomly select 80\% data for training and 20\% for testing, and use the training data to fine-tune the network pre-trained on ShapeNet. Our method performs the best on both ABC~\cite{Koch_2019_CVPR} and Thingi10K~\cite{zhou2016thingi10k}. This demonstrates that our method generalizes well to new datasets, producing minimum average SDE and dealing well with more complex and asymmetric shapes.
	
	\subsection{Computation Time}
\begin{figure*}[t]
		\begin{center}
			\includegraphics[width = 0.8\linewidth]{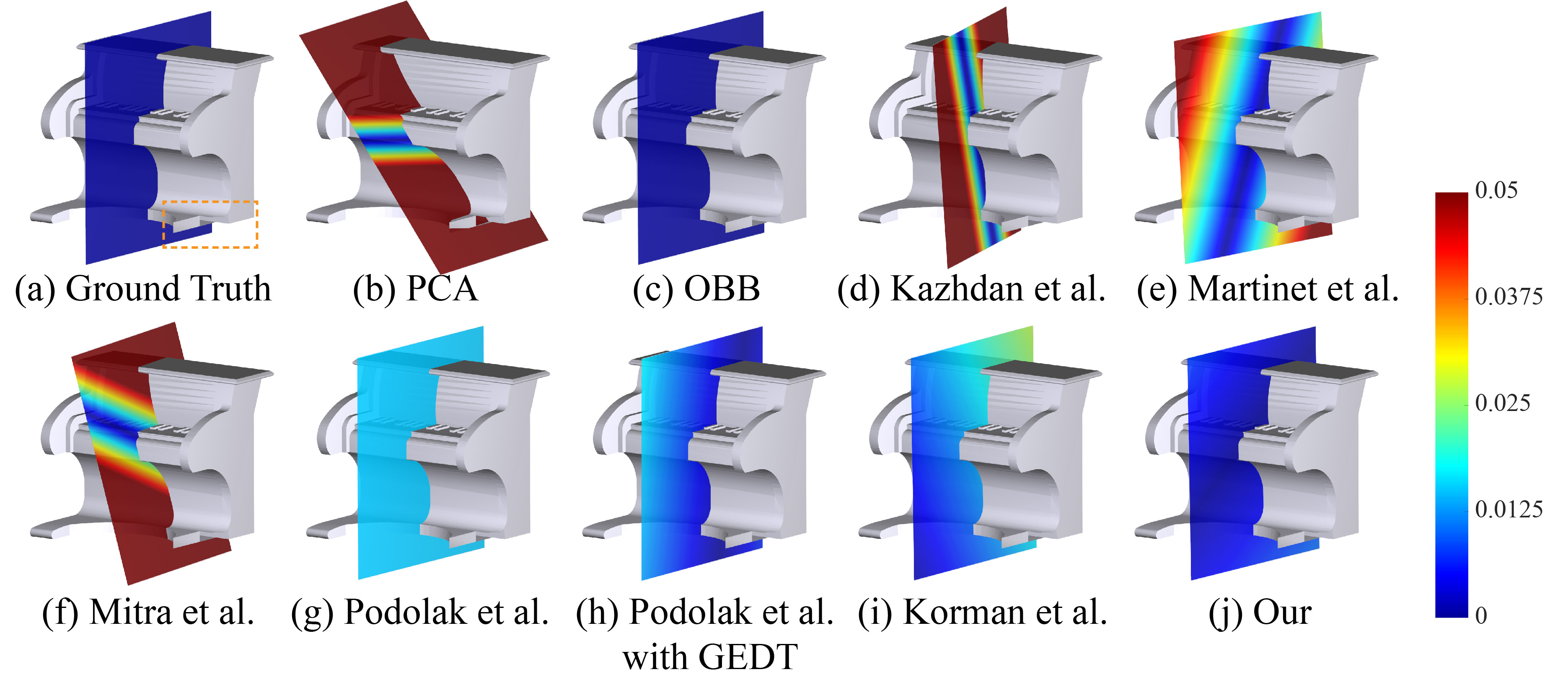}
		\end{center}\vspace{-4mm}
		\caption{Robustness testing by removing parts of the model.  We remove the left leg of the piano and compare the error based on the complete model.
			This demonstrates that our network can also produce accurate results even when the input misses large partial models.}
		\label{partial}
	\end{figure*}

	We calculate the computation time and compare it with alternative methods. Our experiments were carried out on a desktop computer with a 3.6 GHz Intel Core i7-6850K CPU, 128G memory and an NVIDIA TITAN X GPU. For a typical model, such as the piano in Figure \ref{partial} with 1,052 vertices and 4,532 faces, OBB~\cite{chang2011fast}, the method of Kazhdan \textit{et al.}~\cite{kazhdan2002reflective}, the method of Martinet \textit{et al.}~\cite{martinet2006accurate}, the method of Podolak \textit{et al.}~\cite{podolak2006planar}, the method of Podolak \textit{et al.}~\cite{podolak2006planar} with GEDT, the method of Korman \textit{et al.}~\cite{korman2014probably} and the method of Mitra \textit{et al.}~\cite{mitra2006partial} require $0.02$, $0.51$, $2.82$, $3.40$, $5.00$, $0.97$, $0.42$ seconds, whereas our method only needs 1.81 ms, which is hundreds faster than the state-of-the-art methods and achieves real-time performance. This is because these methods use sampling and/or iterative algorithms which increases the computation time.
	In terms of running times, this method is also comparable with PCA
	(typically taking 1.9 ms using CPU) since the trained network is performed on the GPU with powerful computation ability.

	\begin{table*}[t]
		\footnotesize
		\centering
		\caption{
			Comparisons with different methods on partial shape set with large contiguous regions removed. We take the test set and for each shape we randomly choose a radius and a center point, then we remove triangles of the shape that fall inside the sphere. Our method is robust and produces minimum average GTE.
		}\vspace{-3mm}
		\begin{tabular}{lcccccccccc}
			\toprule
			\multirow{2}{*}{Error} & \multirow{2}{*}{PCA} & {Oriented} & Kazhdan \textit{et.al.} &Martinet \textit{et.al.} &Mitra \textit{et.al.} &\multirow{2}{*}{PRST \cite{podolak2006planar}} & PRST \cite{podolak2006planar} & Korman \textit{et.al.}& \multirow{2}{*}{Our}\\
			&&Bounding Box~\cite{chang2011fast}&\cite{kazhdan2002reflective}&\cite{martinet2006accurate}&\cite{mitra2006partial}&&with GEDT&\cite{korman2014probably}&\\
			\midrule
			GTE($\times10^{-2}$) & $9.74$& $1.43$ & $1.92$&$16.8$&$65.1$ &$4.74$& $5.14$&$41.7$& $\mathbf{0.597}$\\
			\bottomrule
		\end{tabular}
		
		\label{partial_table}
	\end{table*}
	
	\begin{figure}[t]
		\centering
		\includegraphics[width=0.7 \linewidth]{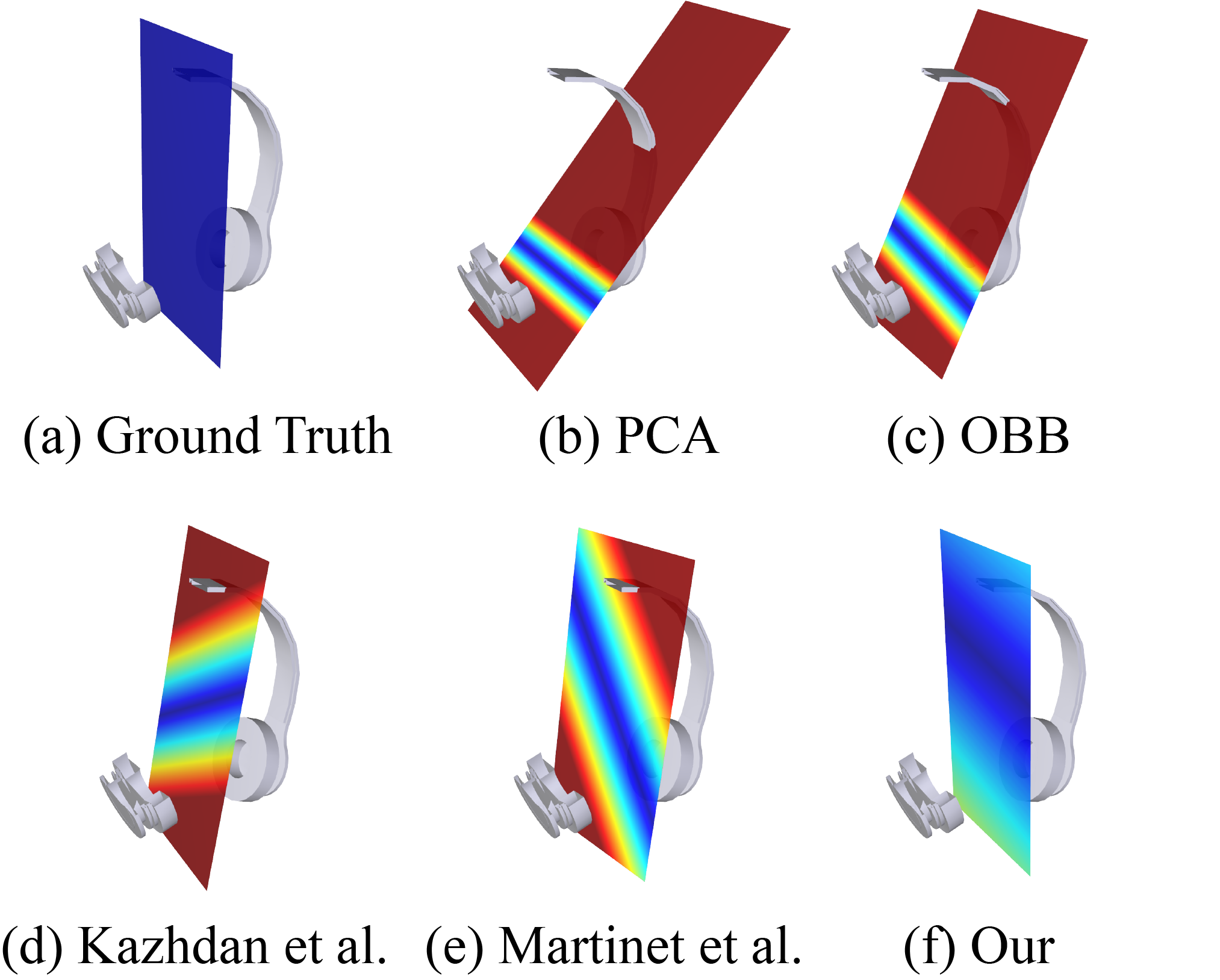}\vspace{-3mm}
		\caption{
			Robustness testing by removing larger parts of the headphone model.
			Our network could produce more accurate result because the network is trained on a large number of shapes including headphones and the feature of partial shape is close to other complete headphones, while other methods lack such knowledge, leading to worse results.}
		\label{robust}\vspace{-4mm}
	\end{figure}

	\subsection{Choice of parameter $w_r$}
	We show an example of visual results with different regularization loss weight in Figure~\ref{fig:para}. We choose $w_r= 25$ for
	training in our paper because it separates multiple symmetry planes properly, unlike Figure~\ref{fig:w1} and Figure~\ref{fig:w10} with small
	$w_r$, and has lower symmetry distance loss than Figure~\ref{fig:w50}.

	\subsection{Application to Shape Completion}
	With the predicted symmetry planes, many geometric tasks would benefit. Here, we apply this to the application of shape completion.
	We show a comparison of shape completion in Figure~\ref{completion}.
	We visualize the Euclidean error between the ground truth part and the generated part of the piano, which is obtained by mirroring the geometry of the left leg along the symmetry plane detected by different methods.
	Our method and OBB give better symmetry planes than other methods, and therefore produce good completed geometry. The method of Martinet \textit{et al.}~\cite{martinet2006accurate} produces worse result because it is designed to detect accurate symmetry. The method of Mitra \textit{et al.}~\cite{mitra2006partial} and Podolak \textit{et al.}~\cite{podolak2006planar} use sampling algorithms to get the reflection planes which is affected by the partial shape. OBB is hardly affected by such missing parts, and our network learns the global shapes of the model through 3D convolutions and numerous training data, and discovers the global symmetry reliably. Thanks to this symmetry detection method, the incomplete shapes with reflective symmetries can be repaired accurately and efficiently.

	\subsection{Robustness}
	In order to test the robustness of our network, we present two different experiments, including noisy and incomplete models. These experiments are motivated by the fact that scanned models often contain noisy and/or incomplete surfaces. As Figure \ref{noise} shows, because most ShapeNet models are non-manifold and some of them have fewer than $1000$ vertices which result in poor noisy models, we first use the method of Huang \etal \cite{huang2018robust} to convert the model to a manifold, before adding Gaussian noise on each vertex along the normal direction.
	It shows that our method produces stable output with smallest error, demonstrating its robustness to small changes of vertex positions.
	
	The second experiment is shown in Figure \ref{partial}, where we remove the left leg surface of the piano and calculate the distance measure based on the original complete model. The distance heatmap shows that our method and oriented bounding box (OBB) are least affected, because our network extracts the global feature through 3D convolutions and pooling, and it learns the global extrinsic shapes of the model. The feature vector of partial piano is close to the complete one because they have very close global shapes. OBB is also insensitive to this situation. The method of Kazhdan \textit{et al.}~\cite{kazhdan2002reflective} has some error because its feature is obtained from voxels which are changed. The method of Martinet \textit{et al.}~\cite{martinet2006accurate} is suitable for accurate symmetry detection, and the incomplete shape affects the result significantly. The methods of Podolak \textit{et al.}~\cite{podolak2006planar} and Mitra \textit{et al.}~\cite{mitra2006partial} use sample points to get the reflection planes, and the distribution of the partial piano points is somewhat different from the complete shape, so the reflection planes have some minor changes. The PCA result is also affected because the shape changes. An example testing robustness against incomplete surfaces is shown in Figure~\ref{robust}. Our network can produce a more accurate result even for the headphone with a substantial part missing, which has the lowest symmetry distance error $1.00\times 10^{-2}$, compared with $2.45\times 10^{-2}$ of PCA, $1.53 \times 10^{-2}$ of OBB, $1.35 \times 10^{-2}$ of the method of Kazhdan \etal~\cite{kazhdan2002reflective}, and $1.30 \times 10^{-2}$ of the method of Martinet \etal~\cite{martinet2006accurate}.
	
\begin{figure}
		\centering
		\includegraphics[width=0.6\linewidth]{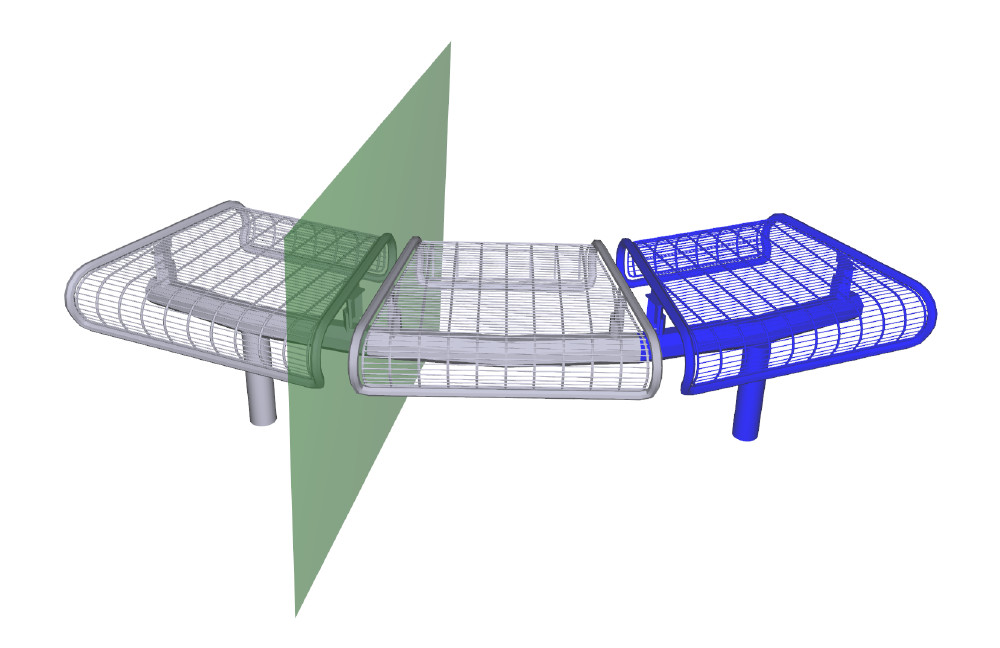}
		\caption{The visual results on the model with extremely large parts removed. The blue areas indicate the parts removed. Our method fails to predict the correct symmetry plane of the original shape.}\vspace{-4mm}
		\label{partial_vis}
	\end{figure}

	We further evaluate on a dataset with large continuous regions removed. We take the test set and for each shape we randomly choose a radius and a center point and remove triangles of the shape that fall inside the sphere. The average GTE is reported in Table \ref{partial_table}.
	Our method is robust and produces minimum average GTE. While voxelization is beneficial, the robustness of our method does not come from it alone, as evidenced by our better performance than the method of Podolak \textit{et al.}~\cite{podolak2006planar} which also uses voxel data. We show results of symmetry detection for partial shapes in Figure ~\ref{partial_vis}, where the blue areas indicate the missing parts.
	Our method cannot predict the correct symmetry plane of this shape because the overall shape and its voxel representation are significantly changed due to the very large missing part.
	
	\begin{table}[h]
		\small
		\centering
		\caption{Comparison of different voxel resolutions. We test the symmetry distance error with resolution from $16^3$ to $128^3$, and the $32^3$ resolution performs best. The error increases with higher resolutions, probably because of overfitting due to limited training data.}\vspace{-2mm}
		\begin{tabular}{lccccc}
			
			\toprule
			Voxel Revolutions & $16^3$ & $32^3$ & $64^3$ & $128^3$\\
			\midrule
			SDE ($\times10^{-4}$) & $1.16$ & $\mathbf{0.861}$ & $1.05$ & $1.28$\\
			\bottomrule
		\end{tabular}
		\label{voxel_revolutions}
	\end{table}

	\subsection{Voxel Resolution and Network Pre-training}\label{sec:res}
	The voxel resolution for the CNN can affect the performance of our network.
	Voxel resolutions from $16^3$ to $128^3$ have been tested and evaluated with the symmetry distance error on the same test set as in Section~\ref{sec:train}.
	In this experiment, we change the number of convolution layers to suit the input voxel size. The number of convolution layers $k_l=log_2(R)$, where $R^3$ is the input voxel size.
	As shown in Table \ref{voxel_revolutions}, the resolution with $32^3$ performs best, which is used as the default resolution. The performance drops with higher resolutions, probably due to the overfitting with large number of parameters.
	
\begin{figure}[t]
		\begin{center}
			\includegraphics[width = \linewidth]{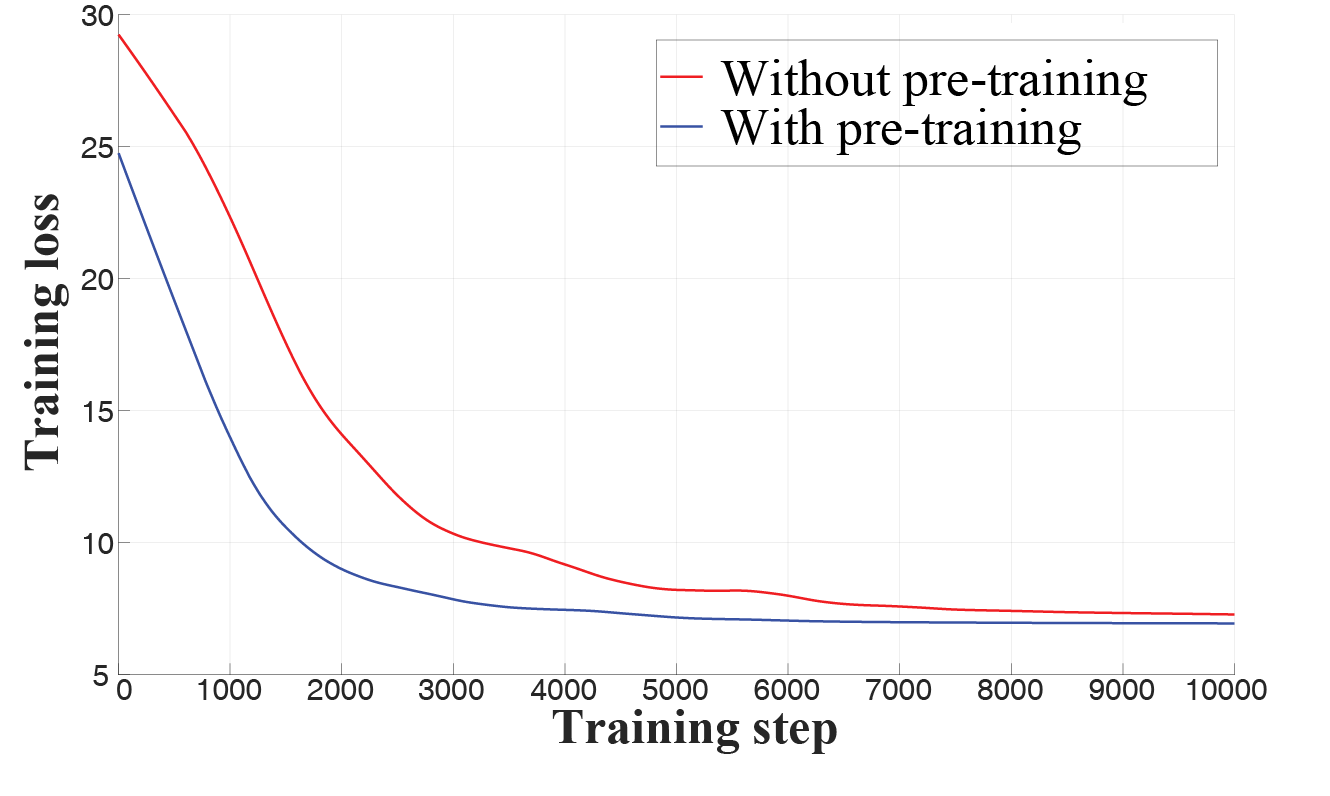}
		\end{center}\vspace{-5mm}
		\caption{{We compare our network with and without pre-trained weights using the output of the method of Kazhdan \textit{et al.}~\cite{kazhdan2002reflective}. We show the training loss ($y$-axis) w.r.t. training steps ($x$-axis).}}\vspace{-4mm}
		\label{pretrain}
	\end{figure}

	Since the method of Kazhdan \textit{et al.}~\cite{kazhdan2002reflective} produces fairly good results, we could use the results to form a supervised loss to train our network and initialize it with these pre-trained weights. As shown in Figure~\ref{pretrain}, we found that the network with pre-training leads to faster convergence and only requires 5000 steps to converge, compared to the network without the pre-training that needs 9000 steps. Although the accuracy on the test data is nearly identical, the initialization with pre-training is helpful for faster convergence.
	
	\begin{figure}[h]
		\centering
		\includegraphics[width=1\linewidth]{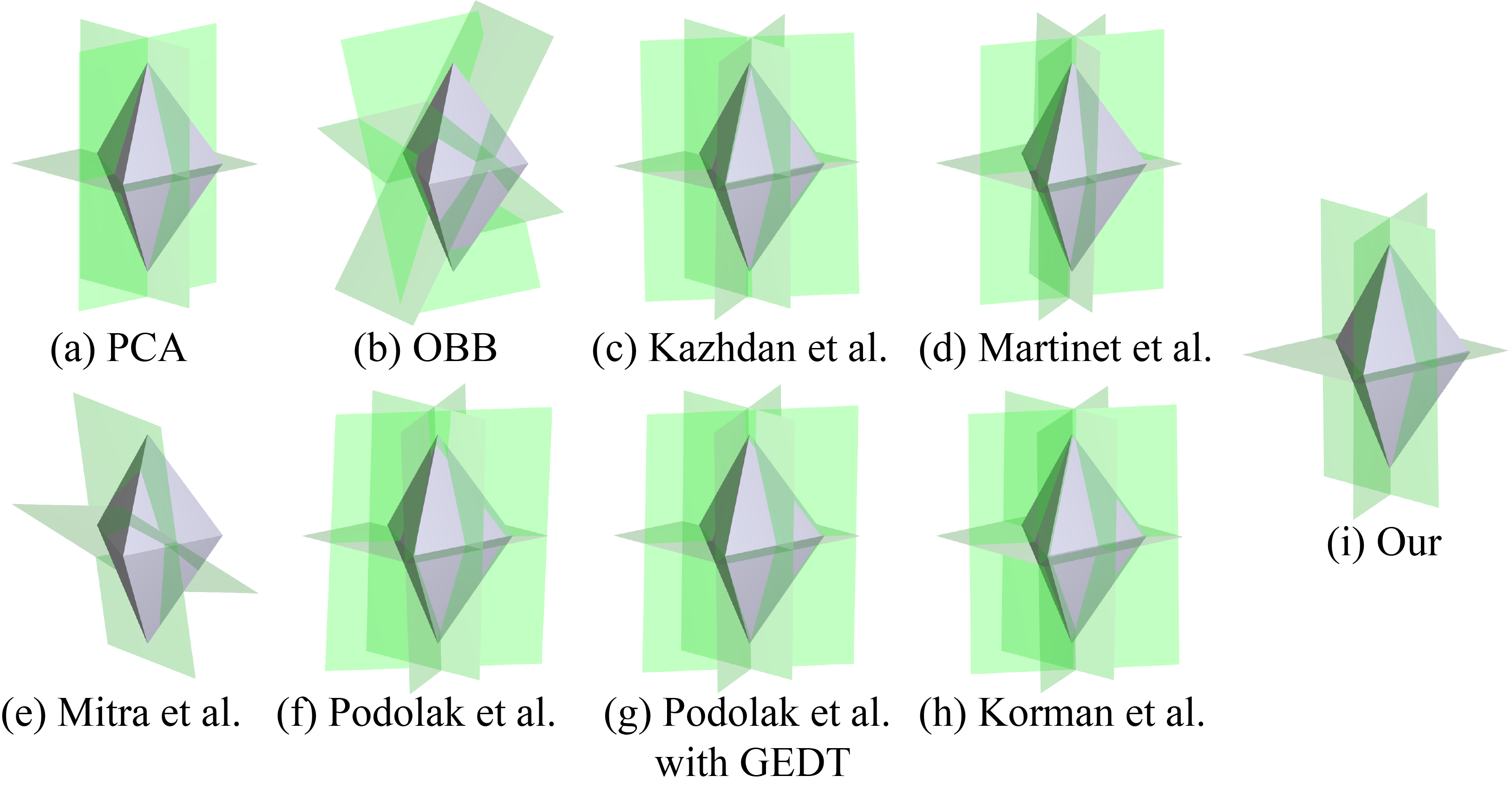}
		\caption{The detection results of different methods on shapes with more than 3 non-orthogonal symmetry planes. Our method detects three non-orthogonal symmetry planes.}
		\label{limitation}
	\end{figure}
	\section{Conclusion and Future Work}
	In this paper, we introduce a novel unsupervised 3D convolutional neural network named PRS-Net, which can discover the planar reflective symmetry of a shape. To achieve this, we develop a symmetry distance loss along with a regularization loss to avoid generating duplicated symmetry planes. We also describe a method to remove invalid and duplicated planes and rotation axes. We demonstrate that our network is robust even when the input has noisy or incomplete surfaces.
	
	Figure~\ref{limitation} shows the detection results of different methods on a shape with more than three, non-orthogonal symmetry planes. Our method detects three non-orthogonal symmetry planes but cannot detect all symmetry planes when the shape has more than three symmetry planes. Although our method could be extended by adding more fully connected layers to predict more symmetry planes, determining the proper number automatically is non-trivial. Our current setting although not perfect is able to handle the majority of cases in practice. In the future, we will investigate \textit{e.g.} reinforcement learning, to predict the number of symmetry planes and their orthogonality, to make the method more general. Moreover, it is interesting to exploit detection of other symmetries including rotational symmetry and intrinsic symmetry using deep learning, which we plan to investigate in the future.

	\ifCLASSOPTIONcompsoc
	\section*{Acknowledgments}
	\else

	\section*{Acknowledgment}
	\fi
	This work was supported by National Natural Science Foundation of China~(No. 61872440 and No. 61828204), Beijing Municipal Natural Science Foundation~(No. L182016),  Royal Society Newton Advanced Fellowship  (No. NAF$\backslash$R2$\backslash$192151), Youth Innovation Promotion Association CAS and Tencent AI Lab Rhino-Bird Focused Research Program (No.JR202024).
	
	\ifCLASSOPTIONcaptionsoff
	\newpage
	\fi

	{\small
		\bibliographystyle{IEEEtran}
		\bibliography{IEEEabrv,reference}
	}
	
\end{document}